# Phase Behavior of Poly(N-isopropylacrylamide) Nanogel dispersions: Temperature Dependent Particle Size and Interactions

J. Brijitta, B.V.R. Tata* and T. Kaliyappan[1]

Materials Science Division, Indira Gandhi Centre for Atomic Research, Kalpakkam 603102, Tamilnadu, India
[1]Department of Chemistry, Pondicherry Engineering College, Pondicherry 605014, India
*Corresponding author: tata@igcar.gov.in

**ABSTRACT**

The phase behavior of poly (N-isopropylacrylamide) nanoparticles dispersed in aqueous medium is investigated as a function of temperature using static and dynamic light scattering techniques. The diameter, $d$ of the particles, as determined by dynamic light scattering measurements on dilute dispersion showed a decrease in size from 273 nm at 25 °C to 114 nm at 40 °C as function of temperature with a sudden collapse of particle volume (volume phase transition) at 32.4 °C. Further this nanoparticle dispersion is found to turn turbid beyond volume phase transition. Static light scattering measurements on samples with intermediate concentration and high concentration showed liquid-like order and crystalline order respectively. The intensity of the Bragg peak of the crystallized sample when monitored as a function of temperature showed crystal to liquid transition at 26.2 °C and a fluid to fluid transition at 31 °C. The occurrence of melting at a volume fraction of 0.85 and the absence of change in number density across the fluid-to-fluid transition suggest that interparticle interaction is repulsive soft-sphere below the volume phase transition. The reported results on the phase behavior of poly(N-isopropylacrylamide) nanogel suspensions are discussed in the light of the present results.

## 1. INTRODUCTION

Under suitable conditions, suspensions of monodisperse organic and inorganic nanoparticles (e.g. polystyrene, polymethyl methacrylate and silica) exhibit liquid, glass, and crystalline phases similar to those observed in atomic systems.[1,2] Sterically as well as charge stabilized nanoparticle dispersions (known popularly as colloidal suspensions) have been intensively studied over the past two decades not only because of the fundamental interest in understanding the co-operative phenomena such as structural ordering, crystallization and glass transition using these systems[1-3] but also for their practical use in a wide range of disciplines, including optical devices, sensors, drug delivery and bio-separations.[4-7] Some of these applications use ordered arrays of these nanoparticles and diverse methods (e.g., self-assembly, sedimentation, template-directed – crystallization, centrifugation, electrophoretic deposition etc.) have been reported for the assembly of nanoparticles into ordered structures.[1, 8-11]

Sterically stabilized nanodispersions, where the particles behave like hard-spheres, can be made to freeze into an face-centered-cubic (fcc) lattice at volume fraction, $\phi$ of 0.5 and into a glassy state at $\phi > 0.56$.[12] In the case of charge-stabilized suspensions, the particles can be frozen into a body-centered-cubic (bcc) or fcc structure depending on the volume fraction of the particles.[13, 14] Though the actual volume fraction at which charge stabilized suspensions freeze into an ordered or disordered (glass) structure is lower than that of hard-sphere suspensions, the corresponding effective hard-sphere volume fractions are comparable.[15] The effective hard-sphere diameter depends on charge on the particle ($Ze$) and the salt concentration ($C_s$) in the suspension but relatively less sensitive to the temperature $T$.[14, 15] Hence one can realize different structural orders in charge stabilized dispersions at ambient conditions by varying suspensions parameters *viz.*, $\phi$, $Ze$ and $C_s$. However, there are dispersions of gel particles, where particle size varies with the temperature, which are popularly known as thermo-responsive macro/nanogel dispersions. Poly(N-isopropylacrylamide) (PNIPAM) is one such hydrogel where the particle size as we all as the interparticle interaction $U(r)$ among the gel particles is tunable by varying the temperature T.[16, 17]

The thermo-responsive PNIPAM hydrogel in its macro and micro/nano forms have been studied extensively as it exhibits remarkable shrinking with increasing temperature.[16-18] A non-continuous collapse transition, known as volume phase transition (VPT), occurs around 34 °C. This transition is tunable to certain extent by random copolymerization of NIPAM with hydrophilic/hydrophobic co-monomers.[19] Since these particles can be synthesized in monodisperse form, they can be assembled into ordered structures with lattice constants in the visible range.[20] Though there have been some studies on ordered nanogels,[17,20,21] there are no systematic studies on the structural ordering and the phase behavior of PNIPAM nanogel aqueous dispersions as function of $T$ and particle concentration $n_p$. Here we report detailed light scattering studies on monodisperse PNIPAM nanogel dispersions with different particle concentrations. For the first time, we report liquid-like order in PNIPAM nanogel dispersion below certain particle concentration, which transforms to gas-like disorder without a change in particle density upon heating. This we identify as fluid-fluid transition and occurs below VPT. At higher concentration, the nanogel dispersion exhibits long-range order, which melts into a liquid-like order upon heating and undergoes fluid-fluid transition (second transition) on further increasing the temperature. These suspensions did not undergo phase separation even when heated beyond VPT. The results are discussed in the light of reported theoretical phase diagrams for PNIPAM microgel dispersions.[17, 22]

## 2. EXPERIMENTAL DETAILS

### 2.1 Synthesis of PNIPAM nanogel particles

The aqueous suspensions of PNIPAM nanogel particles are synthesized by free radical precipitation polymerization as per the procedure described in Ref. 23. 139 mM of N-isopropylacrylamide (purchased from Acros – Belgium), 1.96 mM of N, N'-methylene-bisacrylamide (purchased from Fluka – Germany) and 1.05mM of sodium dodecyl sulfate (SDS) are dissolved in 250 ml of Argon purged water. We have used dust free Milli Q water and AR grade chemicals for the synthesis. The reaction mixture is kept at 70 °C for one hour and then 2.22 mM of potassium persulphate (KPS) is added with vigorous stirring. SDS and KPS were purchased from Rankem- India. The polymerization is carried out at 70 °C for 4 hours under a stream of Argon. Synthesized nanogel suspensions are transferred to cellulose bags (supplied by Hi-media, India) with a molecular weight cut off 10,000 g/mol for dialysis. The purified suspensions were kept in contact with mixed bed of ion exchange resins (Ag501 –X8, Bio-Rad laboratories, Hercules, CA) for further removal of ionic impurities.

### 2.2 Sample Preparation

We have prepared four samples (labeled as S1 to S4) varying in particle concentration $n_p$. A highly concentrated suspension is obtained by centrifuging the mother suspension at 10,000 rpm for one hour. After removing the supernatant, the top and bottom portions of the dense part of the nanogel dispersion are transferred into two light scattering cells. The bottom portion of the dense part (S4) remained as disordered with a bluish hue (Fig. 1). The cell that contains top portion of the dense suspension (S3) when annealed repeatedly from 25 °C to 40 °C showed iridescence (Fig. 1). The particle concentration $n_p$ of samples S4 and S3 are determined to be $10.12 \times 10^{13}$ cm$^{-3}$ and $8.71 \times 10^{13}$ cm$^{-3}$ from static light scattering data. Samples S2 and S1 were prepared by diluting sample S3 by 50% and 0.5 % respectively using Milli-Q water. Samples S2 with $n_p = 4.36 \times 10^{12}$ cm$^{-3}$ appeared slightly turbid and S1 ($n_p = 4.36 \times 10^{11}$ cm$^{-3}$) is transparent (Fig.1). Static and dynamic light scattering studies were carried out on the dilute sample S1 for characterizing the particle size and its polydispersity.

## 2.3 Static and Dynamic Light Scattering Measurements

Static and dynamic light scattering measurements were performed by means of dynamic light scattering set-up (Malvern –UK, 4700 model) consisting up of a goniometer and multi-tau photon correlator and mixed (Ar + Kr) ion laser operating at 514.5 nm wavelength. The multi-tau correlator helps in measuring the normalized intensity autocorrelation function $g^{(2)}(q,t)$ over 11 orders of magnitude in time with the smallest delay time being 50 ns. In a homodyne dynamic light scattering (DLS) experiment $g^{(2)}(q,t)$ is given by

$$g^{(2)}(q,t) = \langle I(q,0)I(q,t) \rangle / \langle I(q,0) \rangle^2 \qquad (1)$$

and is related to the normalized electric field autocorrelation function $f(q, t)$ by the Siegert-relation.[24]

$$g^{(2)}(q,t) = 1 + \beta |f(q,t)|^2 \qquad (2)$$

Here $\beta$ (= ~1) is the coherence factor and the scattering wave vector q = $(4\pi\mu/\lambda)\sin(\theta/2)$. $\mu$ is refractive index of the solvent, $\theta$ is the scattering angle and $\lambda$ is wavelength of the incident laser light.

For a polydisperse sample $f(q,t)$ is related to line-width distribution $G(\Gamma)$ by,

$$f(q,t) = \int_0^\infty G(\Gamma) e^{-\Gamma t} d\Gamma \qquad (3)$$

$G(\Gamma)$ can be calculated from the Laplace inversion of $f(q,t)$. Mean line width $<\Gamma>$ and the relative distribution width $\mu/<\Gamma>$ were obtained by cumulant analysis of $f(q,t)$. The diffusion coefficient $D_0$ is related to $\Gamma$ by $\Gamma = D_0 q^2$. The hydrodynamic diameter $d$ is obtained from Stokes-Einstein relation,

$$d = k_B T / 3\pi\eta D_0 \qquad (4)$$

where $k_B$, $\eta$ and $T$ are the Boltzmann constant, the solvent viscosity and the absolute temperature, respectively. The DLS measurements were carried out at 90°.

In static light scattering (SLS), the angular dependence of the scattered light intensity $I_s(q)$ from the nanoparticle dispersion is measured. The time-averaged scattered intensity measured in vertical-vertical scattering geometry is given as

$$I_s(q) = A P(q) S(q) \qquad (5)$$

where A is a constant[14], $S(q)$ is the interparticle structure factor and $P(q)$ is the particle structure factor.

For a dilute (non-interacting) nanogel dispersion $S(q) = 1$, hence one can obtain radius of gyration $R_g$ of PNIPAM gel particles by analyzing the scattered intensity data at small q and using the relation

$$\frac{1}{I_s(q)} = B(1 + \frac{1}{3}q^2 R_g^2 + \cdots) \qquad (6)$$

The optical diameter[25] $d_o$ of the spherical gel particles is related to $R_g$ by $d_o = 2 (5/3)^{1/2} R_g$.

## 3. RESULTS AND DISCUSSION

### 3.1 Effect of Temperature on Particle Size and Volume Phase Transition

It is known that in colloidal suspensions, the hydrodynamic diameter $d$ obtained from DLS measurements is close to the geometrical diameter if the interactions in the suspension are fully suppressed. This is achieved by preparing a dilute suspension and verifying that the diffusion coefficient measured from DLS measurements and $S(q)$ obtained from SLS measurements (Eq. 5) should be independent of $q$.[26] We have prepared a dilute sample S1 and verified that the sample is non-interacting (gas-like) by performing DLS and SLS measurements. At $T = 25$ °C the average hydrodynamic diameter and size polydispersity are determined to be 273 nm and 0.01, respectively. Further the depolarized scattering intensity measured at different scattering angles is found to be zero. Hence we conclude that PNIPAM particles synthesized by us are monodisperse and spherical in shape. We have varied the temperature of Sample S1 and measured the hydrodynamic diameter as a function of $T$, which is shown in Fig. 2(a). The sudden decrease in $d$ at $T \sim 32.4$ °C corresponds to the volume phase transition (VPT). This transition is found to be reversible upon lowering $T$. These spherical particles display a strong thermo-responsivity, where below a lower critical solution temperature (~ 32 °C)[16] they are highly swollen in water (Fig. 2(b)), but upon increasing the temperature above the lower critical solution temperature they rapidly deswell to a collapsed polymer globule (Fig. 2(b)). Because of this effect, on increasing the temperature to 40 °C from 25 °C, the diameter of the gel particles significantly decrease from 273 nm to 114 nm, which corresponds to about 93% change in the volume of the particles. Further the diameter determined from SLS measurements, known as optical diameter $d_o$, is found to be smaller than $d_h$ below the VPT and almost same above the VPT.[25] These measurements support the earlier conclusions made on the core-shell nature[25, 27] of these particles as depicted schematically in Fig. 2(b).

### 3.2 Liquid-like ordering and Fluid –to-fluid transition

Static light scattering measurements have been carried out on sample S2, which is ten times more concentrated than sample S1, with an aim to characterize the structural ordering. This sample appeared slightly turbid and did not exhibit iridescence even after repeated annealing. The time averaged scattered intensity $I_s(q)$ measured as a function of $q$ showed a peak (see Fig. 3) indicating short-range spatial correlations among the particles in sample S2. The second peak occurs out side the q- range of our instrument, hence not observed. The absence of iridescence and the existence of well defined peak suggest that the sample has liquid-like structural order similar that observed in charged colloidal dispersions.[26]

Since the particle size varies with temperature in PNIPAM gel dispersions, it is interesting to investigate the effect of temperature (in other words the variation in particle size) on the liquid-like order in sample S2. It can be seen from Fig. 3 that first peak height decreases without a shift in first peak position and the peak broadens as $T$ is increased. Further the peak disappears for temperatures at 32 °C and $I_s(q)$ is found to show a slow decrease as $q$ increases. This slow decrease in intensity is due to the variation in $P(q)$ as function of $q$. The absence of a peak in $I_s(q)$ versus $q$ at 32 °C imply a gas-like disorder[26] in sample S2 at this temperature.

In order to identify the liquid-like to gas-like transition in sample S2, we have monitored the first peak intensity $I_{max}$ as function of $T$ and is shown in Fig. 3(b). Interestingly, we observe only a change in slope but not a sudden jump at around 31.5 °C. This suggests that across the transition there is no variation in particle density, $n_p$. Hence we identify this transition as fluid (liquid-like) to fluid (gas-like) transition. The absence of change in $n_p$ across this transition suggests that interparticle interaction is repulsive.

### 3.3 Crystalline order, Melting and Glass-like disorder

Sample S3 with $n_p = 8.71 \times 10^{13}$ cm$^{-3}$ showed iridescence (Fig. 1) due to Bragg diffraction of visible light and also Bragg spot (inset in Fig. 4) for the laser light. These observations suggest that the sample S3 is crystalline and we could record the scattered intensity (Fig. 4) due to Bragg diffraction from crystallites at the center of the light scattering cell, whose (111) plane (we assume the crystal structure to be fcc) are oriented perpendicular to the scattering plane ($xy$ - plane). From the Bragg peak position $q_{111}$ (=2.84 × 10$^5$ cm$^{-1}$) we obtain the particle concentration from the relation

$$n_p = \frac{4}{3\sqrt{2}} \left( \frac{q_{111}}{2\pi} \right)^3 \qquad (7)$$

and the lattice constant from $R_0 = \sqrt{3}(2\pi / q_{111}) = 383$ nm.

In order to identify the melting transition of these PNIPAM gel crystals, we monitor the Bragg peak intensity $I_{max}$ as function of temperature, which is shown in Fig. 5. Notice the sudden decrease in the Bragg peak intensity at 26.2 °C. Slightly above this temperature we found the sample is not iridescent and $I_s(q)$ measured as function $q$ showed a broad peak (inset Fig. 5) implying the liquid-like order in sample S3. Thus we identify 26.2 °C to be the melting temperature of PINPAM nanogel crystals. Further we observed the first peak position to remain same across the meting transition suggesting no measurable change in $n_p$ across this transition. These observations are in close agreement with Wu et al.[17] They determined the melting temperature of their crystallized samples by observing the disappearance of the turbidity peak. Unlike Wu et al[17], we have characterized the structural ordering in our sample to be liquid-like at temperatures beyond the melting temperature.

Upon further increasing the temperature beyond the melting point of PNIPAM nanogel crystals, the peak intensity is found to decrease and showed a change in slope (Fig. 5) at 30.5 °C. Beyond 30.5 °C, the structural ordering in the suspension is found to be gas-like (see inset of Fig. 5) similar to that observed in sample S2. Thus the change in slope observed at 30.5 °C is attributed to the occurrence of fluid (liquid-like order) to fluid (gas-like) transition, as there is no change in particle concentration. Wu et al[17] also have observed fluid to fluid transition at 35 °C, which is beyond the VPT, where as the fluid-to-fluid transition reported here is at 30.5 °C, which is below the VPT. Though the sample appeared slightly turbid beyond 30.5 °C no macroscopic phase separation was observed. The solid to liquid and fluid-to-fluid transition temperatures observed in Sample S3 and sample S2 are consistent with phase diagram reported by Wu and Lai.[22]

Sample S4 with $n_p$ = 10.12 × 10$^{13}$ cm$^{-3}$ appeared bluish in color with out any iridescence under white light illumination. Further, the sample also did not flow upon under tilting the cell upside down. SLS measurements on the sample at 22 °C revealed a broad peak at $q$ = 3.14 × 10$^5$ cm$^{-1}$. These observations suggest that the structural ordering in sample S4 is glass-like. This was further confirmed by observing the non-decay of ensemble averaged electric-field correlation function $f(q,t)$ as a function of time $t$[28], which is shown in Fig. 5. The glass-transition in this sample is being investigated.

### 3.4 Interparticle interactions

At low temperatures (below VPT) the PNIPAM particles are in the swollen state, the maximum content of the particle is only water.[17] Hence the dielectric constant of the particle is almost same as that of the solvent. This results in van der Waals attraction between microgel particles being negligible at low temperatures. Further Wu et al[17] have reported for PNIPAM microgel dispersion that second virial coefficient is positive (i.e. interparticle interaction is repulsive) below VPT and turns negative (interaction is attractive) beyond VPT. So one expects PNIPAM particle to behave essentially as hard-spheres of diameter $d$ up to VPT. If this is the case, the dispersion should have exhibited crystallization in the volume fraction range ~0.5 to 0.74. However the volume fraction (estimated using Eq. 7 and the value of *d*) of the crystallized sample S3 is found to be 0.93 at 25 °C and 0.85 close to the melting. Such high values of volume fraction suggest that PNIPAM particles behave as soft spheres. Further, molecular dynamics simulations with soft-sphere repulsive potential ($U(r) \sim 1/r^n$) with varying softness ($1/n$) showed increase of volume fraction at which crystallization occurs with increase in softness.[29, 30] Hence, we conclude that the interparticle interaction in aqueous PNIPAM nanogel dispersions is soft-sphere repulsive below VPT and is responsible for the phase transitions observed in samples S2 and S3. Beyond VPT attraction might be dominant leading to a phase separation in the form of particle rich phase as predicted by Wu et al.[17] However, we have not observed the phase separation in our samples.

## 4. CONCLUSIONS

We have synthesized monodisperse PNIPAM nanogel particles and investigated the phase behavior of aqueous PNIPAM nanogels suspensions with varying particle concentration using static and dynamic light scattering techniques. The volume phase transition is found to occur at ~ 32.4 °C in a very dilute suspension. For the first time, we have shown that PINPAM particles exhibit liquid-like order below a certain particle concentration and undergo fluid-to-fluid transition at 31.5 °C without a change in the particle number density. At higher particle concentrations, we observe PNIPAM particle freeze into a crystalline state order or glassy state at room temperature. Crystalline order is found to melt at 26.2 °C into a liquid-like order, which undergoes a second transition (fluid- fluid transition) at 30.5 °C. No change in particle number density is observed across the melting transition as well as fluid-to-fluid transition. Though the phase behavior of PNIPAM nanogel dispersions closely resembles that of hard-sphere colloidal system, the volume fraction at which the crystallization appears is much higher. Our observations suggest that below VPT the interparticle interactions are soft-sphere repulsive in nature and responsible for the observed phase behavior. Wu et al[17] predicted phase separation driven by van der Waals attractions at temperatures beyond VPT. This prediction needs to be verified through further experiments.

## ACKNOWLEDGEMENTS

We thank Dr. A.K. Arora, Dr. C.S. Sundar for useful discussions and Dr. Baldev Raj for support and encouragement. First author acknowledges UGC-DAE CSR for financial support.

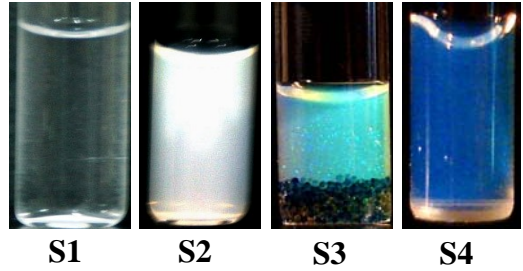

|  S1 | S2 | S3 | S4 |

**Fig 1:** Photographs of samples S1, S2, S3 and S4 with varying particle concentration. Sample S3 shows iridescence (crystalline) for visible light and mixed bed of ion exchange resins can be seen at the bottom of the cell. Sample S4 is bluish in color without iridescence.

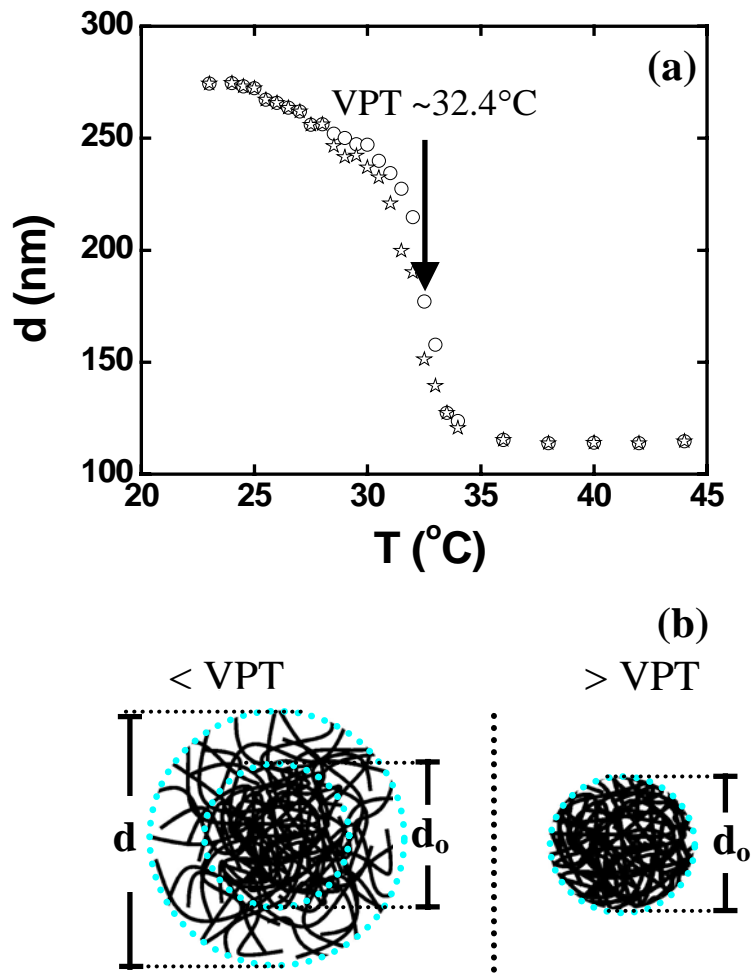

**Fig 2:** **(a)** Hydrodynamic radius $d$ as function of temperature T in increasing and decreasing cycles measured in sample S1. The arrow indicates the VPT temperature. **(b)** Schematic representation of core-shell nature of PNIPAM nanogel particles below and above VPT; $d_o$ is the optical diameter of the particles as determined from static light scattering.

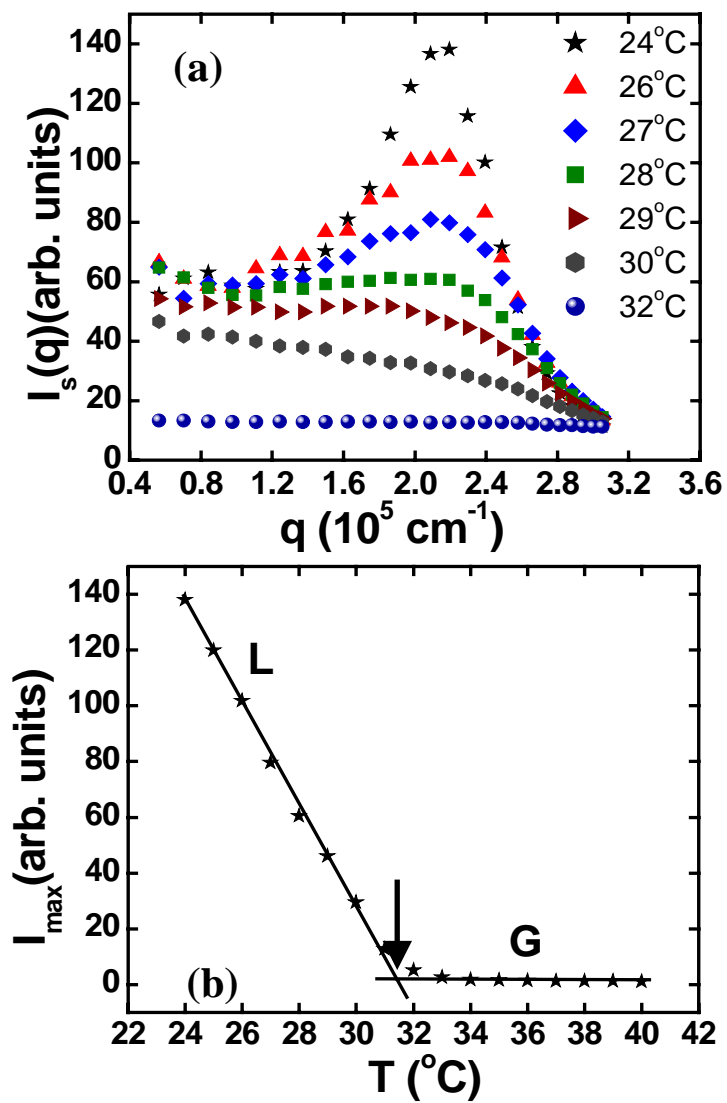

**Fig 3: (a)** Scattered intensity $I_s(q)$ as a function of $q$ for sample S2 at different temperatures. **(b)** Peak intensity $I_{max}(q)$ as a function of $T$ for sample S2. Arrow indicates the transition temperature (~31.5 °C) of liquid-like order (L) to gas-like (G) disorder in sample S2.

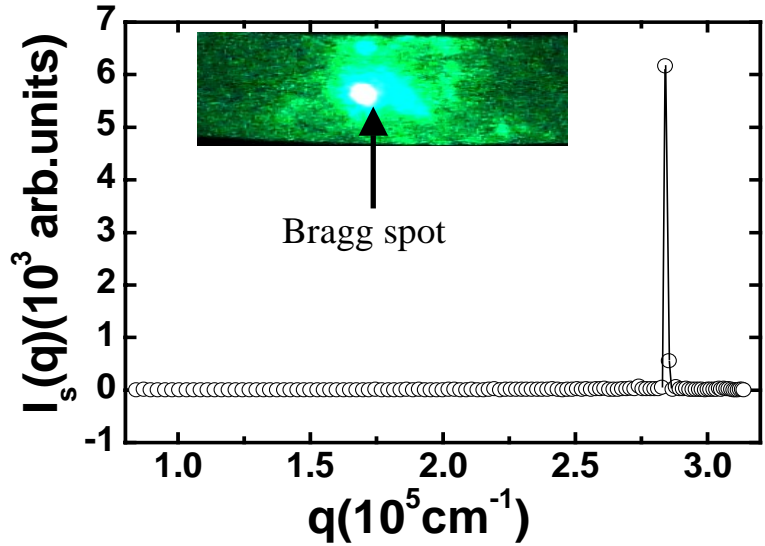

**Fig 4:** $I_s(q)$ as a function of $q$ for sampe S3 at T=22 °C. The inset shows the photograph of the Bragg spot (arrow) from crystallites in sample S3, when illuminated by 514.5 nm wavelength of Ar-Kr mixed ion laser.

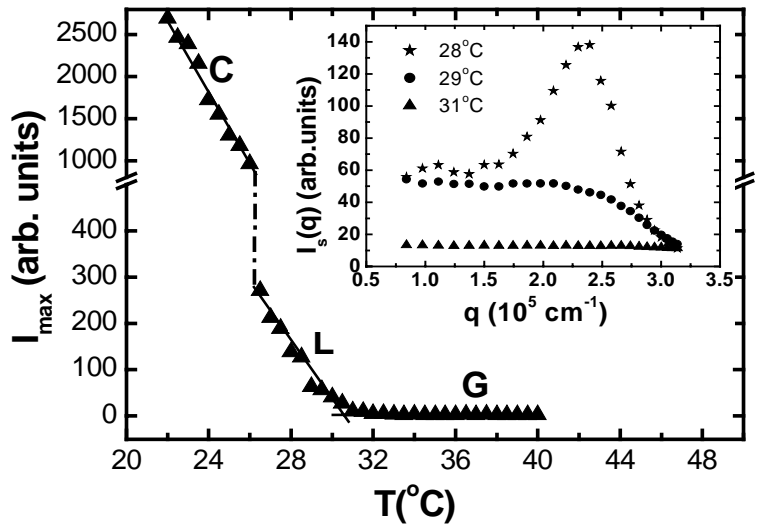

**Fig 5:** Bragg peak intensity $I_{max}(q)$ as a function of $T$ for sample S3. The inset shows $I_s(q)$ vs. $q$ for sample S3 at temperatures beyond the melting temperature of crystallites in Sample S3. C, L, and G represent the temperature region where sample S3 exhibited crystalline, liquid-like and gas-like disorder, respectively

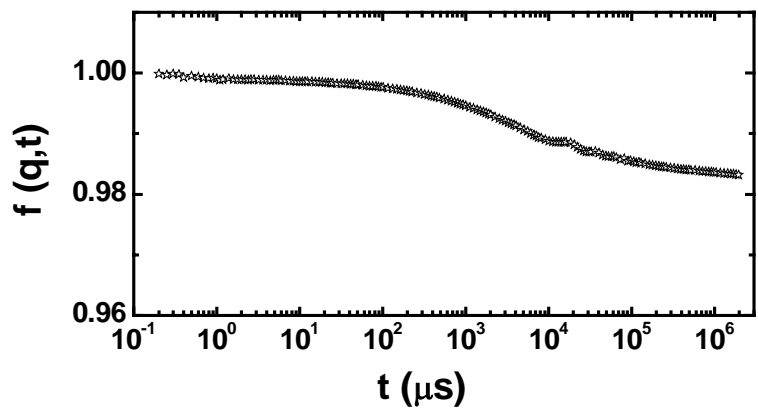

**Fig 6:** *f(q,t)* as a function of *t* for sample S4 measured at the peak position ($q = 3.14 \times 10^5$ cm$^{-1}$) and at T=25 °C.